\documentclass[twocolumn,10pt,superscriptaddress,showpacs]{revtex4}
\usepackage{anysize}
\usepackage{algorithmicx,algorithm}
\usepackage{amsfonts}
\usepackage{amssymb,amsmath}
\usepackage{booktabs} 
\usepackage{color}
\usepackage{epsf}
\usepackage{epsfig}
\usepackage{float}
\usepackage{graphicx}
\usepackage{hyperref}
\usepackage{mathrsfs}
\usepackage{microtype}
\usepackage{subfigure}
\usepackage{siunitx}
\usepackage{ulem}
\usepackage[usenames,dvipsnames]{xcolor}
\usepackage{algpseudocode}

\begin{document}
\title{End-to-End Quantum Machine Learning Implemented with Controlled Quantum Dynamics}

\author{Re-Bing Wu}\email{rbwu@tsinghua.edu.cn}
\affiliation{Department of Automation, Tsinghua University, Beijing, 100084, China}
\affiliation{Beijing National Research Center for Information Science and Technology, Beijing, 100084, China}

\author{Xi Cao}
\affiliation{Department of Automation, Tsinghua University, Beijing, 100084, China}
\author{Pinchen Xie}
\affiliation{Program in Applied and Computational Mathematics, Princeton University, Princeton, NJ 08544, USA}
\author{Yu-xi Liu}
\affiliation{Institute of Microelectronics, Tsinghua University, Beijing, 100084, China}

\date{\today}
\begin{abstract}
Toward quantum machine learning deployed on imperfect near-term intermediate-scale quantum (NISQ) processors, the entire physical implementation of should include as less as possible hand-designed modules with only a few ad-hoc parameters to be determined. This work presents such a hardware-friendly end-to-end quantum machine learning scheme that can be implemented with imperfect near-term intermediate-scale quantum (NISQ) processors. The proposal transforms the machine learning task to the optimization of controlled quantum dynamics, in which the learning model is parameterized by experimentally tunable control variables. Our design also enables automated feature selection by encoding the raw input to quantum states through agent control variables. Comparing with the gate-based parameterized quantum circuits, the proposed end-to-end quantum learning model is easy to implement as there are only few ad-hoc parameters to be determined. Numerical simulations on the benchmarking MNIST dataset demonstrate that the model can achieve high performance using only 3-5 qubits without downsizing the dataset, which shows great potential for accomplishing large-scale real-world learning tasks on NISQ processors.
\end{abstract}


\maketitle

\section{Introduction}
Quantum Computing has entered the NISQ (Noisy Intermediate-Scale Quantum) era \cite{Preskill2018} in which it may surpass classical computing with even imperfect quantum hardware \cite{Arute2019}. As one of the most promising applications, quantum machine learning is drawing intense attention \cite{Biamonte2017,Dunjko2018} for its potential supremacy on solving large-scale real-world learning tasks with quantum computers. With ideal programmable and error-tolerant quantum computers, many quantum subroutines such as Quantum Fourier Transform or Grover search can be applied to speed up the training or inference process,  e.g., quantum supporting vector machine for classification problems \cite{Rebentrost2014}, quantum principal component analysis \cite{Lloyd2014} and quantum generative adversarial learning \cite{Lloyd2018a,Hu2019}.

To enable quantum machine learning algorithms on NISQ processors, a realistic approach is to construct quantum neural-network (NN) models with parameterized quantum circuits (PQC) \cite{Zhu2019,Benedetti2019} that is trained by classical optimization algorithms. Such hybrid quantum-classical models have universal approximation capabilities and are able to achieve classically intractable feature learning tasks \cite{Lloyd2018}. Various applications have been put forward for quantum simulation \cite{Kandala2017}, combinatorial optimization \cite{Farhi2017} and machine learning problems \cite{Farhi2018}. Moreover, feature selection schemes were also proposed, such as the quantum kitchen sink \cite{Wilson2018} and quantum metric learning \cite{Lloyd2020}, to efficiently encode the data into quantum states by learning the representation.

The PQC-based machine learning has also been experimentally demonstrated by shallow circuits on NISQ processors (with no greater than 20 qubits) \cite{Benedetti2019}) on classification \cite{Havlicek2019}, clustering \cite{Otterbach2017} and generative \cite{Hamilton2019} learning tasks. To our knowledge, few of them were tested with real-world datasets, and the achieved performance was still poor unless when the dataset is downsized. For example, the precision of recognizing hand-written digits in the benchmarking MNIST dataset \cite{LeCun1998} is no higher than what can be achieved by a simple classical logistic regression model, and in most cases the original images have to be down-sampled to make compromises with scarce quantum resources (e.g. limited number of qubits and decoherence time) \cite{Grant2018,Farhi2018,Huggins2019,Kerenidis2018,Wilson2018,Sweke2019,Henderson2020}.

On top of limited quantum resources, the architecture of current PQC ansatz also challenges the development of practical NISQ processors, because the designed gate sequences cannot be trivially implemented due to the incompatibility of the PQC topology with the available quantum processors with sparse qubit-qubit connectivity. For example, non-local gates may need to be realized through a series of intermediate operations (e.g., SWAP) due to the lack of interactions between target qubits, while local gates may not be directly implementable when the qubit-qubit coupling cannot be freely turned off. Hence, the designed circuit must be properly mapped to actual interconnect topology of quantum chip, and consequently the compiled quantum circuit is usually deeper and more complicated.

All these demands call for a hardware-friendly quantum machine learning scheme that can be efficiently deployed on NISQ processors with full consideration of their on-chip interconnect topologies. The entire physical implementation should include as less as possible hand-designed modules with only few {ad-hoc} parameters to be determined, so that as less as possible errors are introduced to the quantum machine learning process. In other words, the scheme should provide an end-to-end data pipeline that automatically extracts the features for inference from the input data, so that the overall performance can be better improved. This has become an influential trend in classical deep learning, especially in big data applications~\cite{LeCun2015}.

In this paper, we propose that such end-to-end quantum learning model can be naturally realized by dynamical quantum evolution manipulated by the laboratory hardware control devices. In the following, we will show in Sec.~\ref{sec:model} how this can be done by re-parameterizing the PQC with control variables and by introducing a data-to-control interface for automatic feature selection. Then, in Sec.~\ref{sec:training} we provide the hybrid quantum-classical training algorithm, following which simulation examples with 3-5 qubits are given to demonstrate the effectiveness of the proposed end-to-end learning scheme in Sec.~\ref{sec:simulation}. Finally, concluding remarks are made in Sec.~\ref{sec:conclusion}.

\begin{figure}
\centering
\includegraphics[width=0.95\columnwidth]{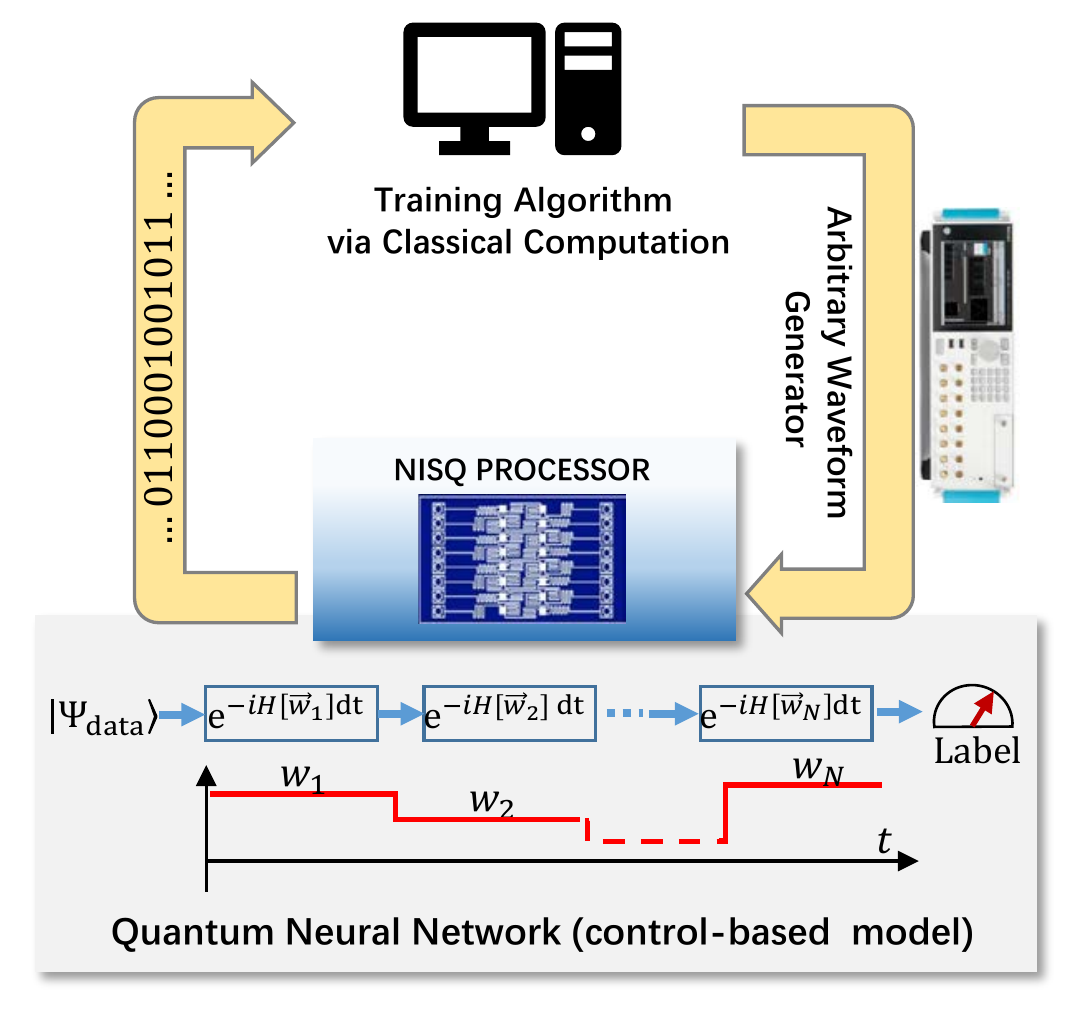}
\caption{The physical control associated with a hybrid quantum-classical algorithm deployed on a NISQ processor. The classical computer runs the training algorithm to instruct the arbitrary waveform generator (AWG) to iteratively update the control pulses according to the qubit measurement outcomes. The controlled quantum evolution forms an equivalent quantum NN parameterized by time-dependent AWG parameters.}
\label{fig:flowchart}
\end{figure}

\section{Quantum End-to-End Learning Model}\label{sec:model}
To illustrate how the end-to-end quantum learning can be implemented by controlled quantum dynamics, let us start from a classification learning task with a set of $Z$ training samples $(x^{(k)},y^{(k)})$, where $x^{(k)}\in\mathbb{R}^d$ is the input data represented by $d$-dimensional vectors and $y^{(k)}\in\{1,\cdots,r\}$ is the corresponding label. For most quantum machine learning models, the input $x$ is first transferred to the quantum state $|\Psi(x)\rangle$ of the register through an encoder circuit. After being processed by a succeeding quantum circuit represented by a parameterized unitary transformation $U({\bf w})$, where ${\bf w}$ is the hyper-parameters (e.g., rotating angles of $X$, $Y$ or $Z$ gates) of the circuit. The output state is measured under a POVM measurement $\{M_1,\cdots,M_r\}$, in which the operator $M_k$ is associated with the $k$th class to be discriminated. The conditional  probability of obtaining $y$ for a given input $x$ and circuit $U({\bf w})$ is then $P(y|x,{\bf w}) = \langle\Psi(x)|U^\dag({\bf w})M_{y}U({\bf w})| \Psi(x)\rangle$, based on which we can define the empirical loss:
\begin{equation}\label{eq:empirical loss}
  L[{\bf w}]=1-Z^{-1}\sum_{k=1}^Z P(y^{(k)}|x^{(k)},{\bf w}).
\end{equation}

The commonly applied PQC model for machine learning usually consists of layered parameterized one-qubit or two-qubit quantum gates. As is discussed above, the assigned gates in such black model may not be directly implementable due to the limited qubit-qubit connectivity, and hence a more hardware friendly scheme is desired.

\subsection{From circuit model to control model}
In practice, the PQCs performed on quantum chips are always realized through a set of hardware control and measurement devices. For example, in the experimental superconducting quantum computing system~\cite{Johnson2011,Barends2013,Gu2017} shown in Fig.~\ref{fig:flowchart}, the entire PQC is dictated by the control pulses produced from an arbitrary waveform generator (AWG) for implementing the individual designed quantum gates, and the inference is made by readout the qubit states. Therefore, it is natural to replace by these control amplitudes the gate parameters in the unitary transformation $U({\bf w})$ realized by the PQC. An obvious and significant advantage of such control-based model is the hardware friendliness because the control parameters are directly tunable in experiments.

The control-based model can also be treated as a layered quantum feedforward neural network represented by its time-evolving quantum dynamics steered by the Schr\"{o}dinger equation:
\begin{equation}\label{}
  |\dot{\Psi}(t)\rangle = -i\left[H_0+\sum_{\ell=1}^m w_\ell(t)H_\ell\right]|\Psi(t)\rangle,
\end{equation}
where $|\Psi(t)\rangle$ is the quantum state (initially prepared at $|\Psi(t_0)\rangle=|0\rangle$) of the entire system. When the AWG pulses $w_1(t),\cdots,w_m(t)$ consists of $M$ piecewise-constant sub-pulses over $M$ sampling periods, the states $|\Psi(t_k)\rangle$ ($k=1,2,\cdots,M$) at the end of each sub-interval form the layers of the quantum NN. The interconnection between these layers are realized by the unitary evolution operators over these sub-intervals, which are parameterized by control variables $
\vec{w}_k=[w_1(t_k),\cdots,w_M(t_k)]$. The equivalent depth of the quantum NN is thus the number of AWG sampling periods during the entire quantum evolution, while the width is determined by the number of qubits.

The control-based model is a generalization of the gate-based PQC model because any gate operation must be eventually realized through physical control pulses. In special cases when the control Hamiltonians are mutually commutable, it is equivalent with a PQC model, because each control parameter governs a parameterized gate. However, under more general circumstances with limited tuning ability of qubit-qubit couplings, the compilation of gate-based PQCs becomes much more complicated, but the control-based scheme can easily adapt to the on-chip interconnect topology without having to artificially split the model into separate gates.

\subsection{From hand-designed to auto-selected features}
\begin{figure}
\centering
\includegraphics[width=1\columnwidth]{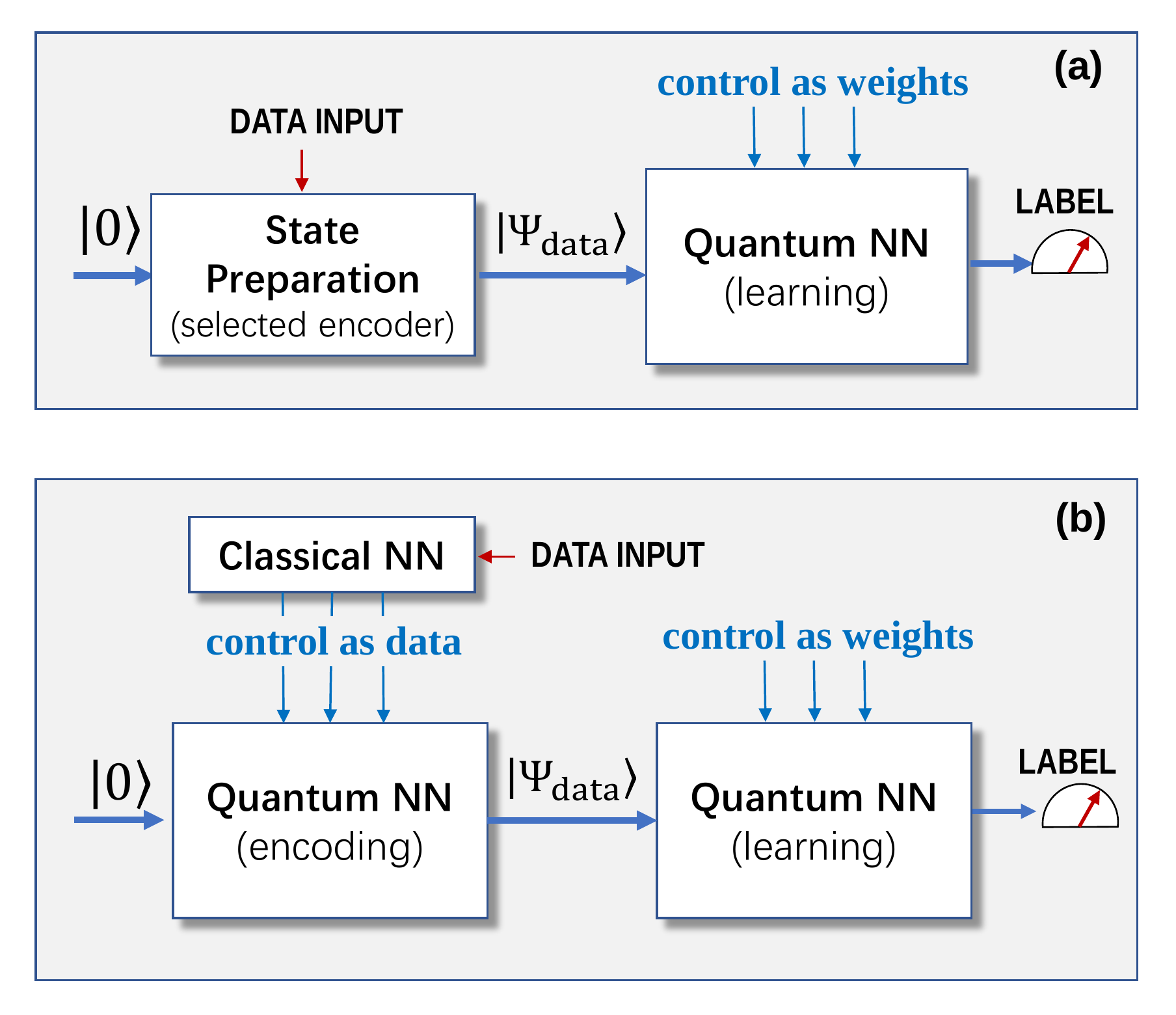}
\caption{Quantum control-based end-to-end learning models. (a) The data is encoded to the quantum state through a pre-selected encoder; (b) the data is encoded to the quantum state through a data-to-control interface (a classical NN) via a selected set of agent control variables. Following the encoding control processes, inference control pulses are applied as weights to infer the class that the input belongs to.}
\label{fig:QE2E}
\end{figure}

In most PQC-based learning models, the data vector is mapped to the quantum state using a pre-selected encoder to represent the set of hand-designed features. As schematically shown in Fig.~\ref{fig:QE2E}(a), the encoder first `translates' the data vector to a quantum state, and then applies a control pulse to physically prepare the system in this state. The complexity of the control design depends on the encoded state, which could be very expensive when the state is highly entangled (e.g., in the amplitude encoding scheme for exploiting the superposition of quantum states). The scheme also becomes impractical when dealing with large-size datasets because every single sample needs an individually designed control pulse.

We propose that the `translation' from the data vector to the quantum state can be designed in an {\it implicit} and {\it automatic} manner. As is shown in Fig.~\ref{fig:QE2E}(b), a data-to-control interface (e.g., a classical NN) is introduced to transform the data vector into a selected set of agent control variables, which then conveys the information about the received data to the quantum state they steer to. The encoded state is not explicitly (and nor necessarily) known unless being reconstructed through quantum state tomography.

The introduced data-to-control interface can be taken as a hidden NN layer that feeds the classical data into the quantum NN. It can be trained together with the rest part of the quantum NN. Once the interface is determined, the encoding control pulse will be automatically generated in response to the input sample.

It should be noted that the encoding scheme also brings favored nonlinearity through the nonlinear control-to-state mapping, which is crucial for improving the model expressivity in complex learning tasks. Later we will show in the simulation examples that the nonlinearity in control-to-state mapping plays a crucial role in achieving high learning performance.

\section{The Training process}\label{sec:training}
Now we have build up a machine learning model based on the controlled quantum dynamics. The training of the learning model can thus be naturally transformed to a quantum optimal control problem \cite{Glaser2015}. Suppose that the involved control pulses contains $M$ sampling periods. We assign the control variables
  $w^{\rm code}=\{\vec{w}_1,\cdots,\vec{w}_{M_0}\}$
in the first $M_0$ sampling periods for data encoding, and $w^{\rm infer}=\{\vec{w}_{M_0+1},\cdots,\vec{w}_{M}\}$ in the rest sampling periods for the inference.

The data-to-control interface can be chosen as arbitrary linear or nonlinear function. For illustration, we select a perceptron layer, i.e., each element of $w^{\rm code}$ is
\begin{equation}\label{eq:perceptron}
w_i^{\rm code} = B\cdot \frac{e^{\sum_j W_{ij}x_j+b_i}-1}{e^{\sum_j W_{ij}x_j+b_i}+1},
\end{equation}
where $x_j$ is the $j$th element of the input $x\in \mathbb{R}^d$, $W=\{W_{ij}\}$ and $b=\{b_i\}$ are the weight matrix and bias vector of the perceptrons, and $B>0$ is the bound of the control amplitudes. Because the bias term $b$ can be merged into $W$ by extending $x$ to $(x^T,1)^T$ and $W$ to $(W,b)$, we will ignore $b$ for simplicity. Thus, the hyper-parameters to be trained are ${\bf w} = (W,w^{\rm infer})$.

The model is trained by minimizing the empirical loss $L[{\bf w}]$ defined by Eq.~(\ref{eq:empirical loss}). Similar to most hybrid quantum-classical algorithms, these hyper-parameters are to be tuned along gradient-descent directions of $L[{\bf w}]$. Since the gradient vector is not directly computable on the NISQ processors, we need to sequentially perturb each hyper-parameter, evaluate the change of empirical loss via ensemble measurements on $L({\bf w})$ and estimate its gradient with respect to ${\bf w}=(W,w^{\rm infer})$ via the finite difference:
\begin{equation}\label{}
  \frac{\partial L}{\partial {\bf w}_j}\approx \frac{L({\bf w}+\Delta\cdot {\bf e}_j)-L({\bf w})}{\Delta},
\end{equation}
where ${\bf e}_j$ is the unit vector along which the $j$th element of ${\bf w}$ is perturbed by $\Delta$.

Let $n$ be the number of qubits, and each qubit is manipulated by $c$ independent control fields (e.g., the bias field for frequency tuning or Rabi driving fields for flipping qubits). Then there are $N_{\rm code}=cnM_0$ encoding control variables to be generated by $(d+1)N_{\rm code}$ weight variables in $W$ and $N_{\rm infer}=cn(M-M_0)$ inference control variables to be directly tuned. This implies that, to evaluate the gradient with a given input sample, about $(d+1)N_{\rm code}+N_{\rm infer}$ ensemble measurements will be required on the conditional probability $P(y|x,{\bf w})$. The experimental overhead can easily exceed the ability of current NISQ processors for large-size and high-dimensional datasets.

Nonetheless, observing that the gradient of $P(y|x,{\bf w})$ with respect to the entries of $W$ can be decomposed (via the chain rule) as:
\begin{eqnarray}\label{eq:dL/dW}
  \frac{\partial P(y|x,{\bf w})}{\partial W_{ij}} 
& =&  \frac{\partial P(y|x,{\bf w})}{\partial w^{\rm code}_i}\frac{B^2-(w^{\rm code}_i)^2}{2B}x_j,
\end{eqnarray}
where $1\leq i\leq N_{\rm code}$ and $1\leq j\leq d+1$, we only need to experimentally measure $\frac{\partial P(y|x,{\bf w})}{\partial w^{\rm code}_i}$, with the rest parts handled by a classical computer. In this way, the experimental burden can be greatly relieved because the number of required ensemble experiments is reduced from $(d+1)N_{\rm code}+N_{\rm infer}$ to $N_{\rm code}+N_{\rm infer}$ (the total number of control variables), which is not explicitly dependent on the dimensionality of the data space.

Based on the measured gradient, we can apply the widely used stochastic gradient algorithms for machine learning, which had been demonstrated to be powerful in robust quantum control \cite{Wu2019} and quantum approximate optimization algorithms \cite{Dong2019}. Roughly speaking, in each iteration we randomly select a small batch of samples, apply the encoding and inference control fields, and measure the conditional probability $P(y|x,{\bf w})$ and its gradient for each sample. The averaged gradient over these samples is then used to update the model hyperparameters ${\bf w} = (W,w_{\rm infer})$. The detailed pseudo-code of the training algorithm can be found in Algorithm \ref{alg:qEEL}.

\begin{algorithm}[tb]
   \caption{Quantum End-to-End Learning}
   \label{alg:qEEL}
\begin{algorithmic}
   \State {\bfseries Input:} training dataset $\{x^{(k)},y^{(k)}\}$, the mini-batch size $m$, and the learning rate $\alpha$.
   \State {\bfseries Output:} the hyper-parameter ${\bf w}=(W,w^{\rm infer})$.
   \Repeat
   \State Sample a mini-batch from the training dataset.
   \For{$i=1$ {\bfseries to} $m$}
   \State Feed the $i$th sample in the mini-batch, say $(\bar{x},\bar{y})$, to generate the encoding control variables $w^{\rm code}$ via $W$.
   \State Synthesize the full control pulse with current $w_{\rm code}$ and $w_{\rm infer}$.
   \State Perturb the control variables and measure their gradients $\frac{\partial P(\bar{y}|\bar{x},{\bf w})}{\partial w^{\rm code}}$ and $\frac{\partial P(\bar{y}|\bar{x},{\bf w})}{\partial w^{\rm infer}}$, and use Eq.~(\ref{eq:dL/dW}) to evaluate $\frac{\partial P(\bar{y}|\bar{x},{\bf w})}{\partial W}$ from $\frac{\partial P(\bar{y}|\bar{x},{\bf w})}{\partial w^{\rm code}}$.
   \EndFor
   \State Compute the average gradient $\frac{\partial L[{\bf w}]}{\partial W}$ and $\frac{\partial L[{\bf w}]}{\partial w^{\rm infer}}$ over the selected mini-batch, and update ${\bf w}$ with a selected optimizer (e.g., Adam).
   \Until{Empirical loss is sufficiently small.}
\end{algorithmic}
\end{algorithm}

\section{Simulation results}\label{sec:simulation}
Now we apply the proposed end-to-end learning model to the MNIST dataset for recognition of handwritten digits. To demonstrate the effectiveness and efficiency, we use a simple chain system of $n=3\sim 5$ qubits, which is typical in solid-state quantum computing, as the physical realization of the NISQ processor. The Hamiltonian reads:
\begin{equation}\label{eq:hamiltonian}
  H(t) = \sum_{1\leq i< n}g_{ij}\sigma_z^i\sigma_z^{i+1} +\sum_{k=1}^n \left[w_{kx}(t)\sigma_x^k+w_{ky}(t)\sigma_y^k\right]
\end{equation}
where $\sigma_{\alpha}^k$ ($\alpha=x,y,z$) are standard Pauli matrices for the qubits. The neighboring qubit-qubit coupling strengths are $g_{12}=1.5$MHz, $g_{23}=2.0$MHz, $g_{34}=2.5$MHz, and $g_{45}=3.0$MHz, respectively. These qubits are addressed by control fields $w_{kx}(t)$ and $w_{ky}(t)$ along $x$- and $y$-axis, respectively. In all simulations, we fix the AWG sampling periods as $5$ns and set the control bounds $B=25$MHz.

\subsection{The training process}

To train the learning model, we use 46993 samples associated to 8 digits $\{0,2,3,4,5,6,8,9\}$ (because 3-qubit models can discriminate at most 8 digits). The POVM measurement for inference is chosen to be $\{M_j= |j\rangle\langle j|,~j=000,001,\cdots,111\}$ under the $\sigma_z$-basis of the first three qubits. The original $28\times28$-pixel sample images are converted to $d=28\times28+1=785$ dimensional vectors after merging the bias vector $b$ into $W$.

\begin{figure}
	\centering
	\includegraphics[width=1.0\columnwidth]{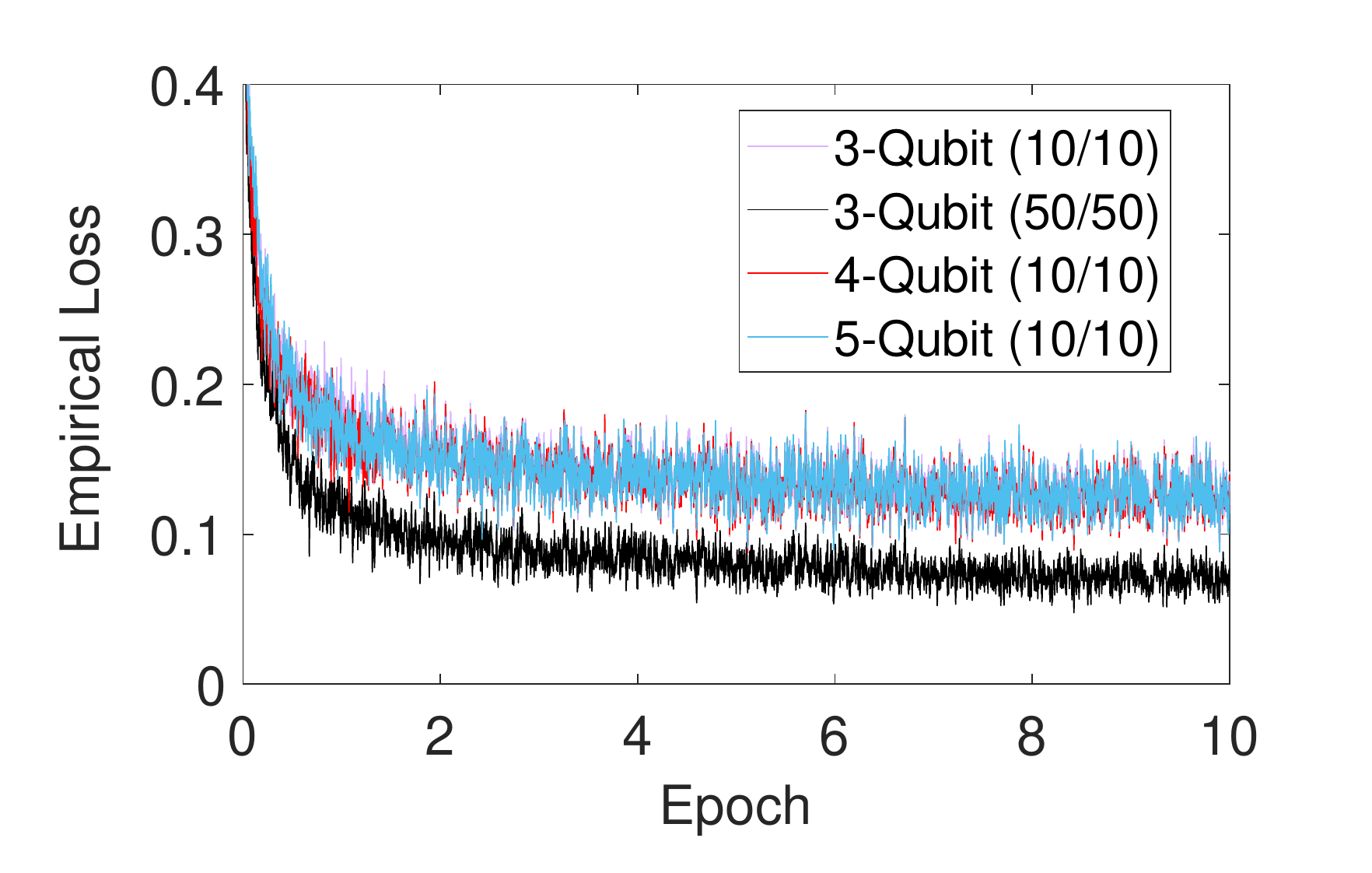}
	\caption{The learning curves for the training of quantum end-to-end learning models. The empirical loss evaluated over each mini-batch (smoothed over the past 100 iterations) during the training process for models with $3\sim 5$ qubits and various numbers of encoding/inference layers.}
	\label{fig:training}
\end{figure}

We first train learning models with fixed depth (all using 10 encoding layers and 10 inference layers) and vary the number of qubits from 3 to 5, and an additional 3-qubit model with 50 coding layers and 50 inference layers for comparison. From the learning curves shown in Fig.~\ref{fig:training}, the deeper model learns remarkably better, in which the empirical loss can be reduced to below $10\%$ after a few epochs (an epoch means that all training samples are traversed for once). In contrast, the performance remains almost unimproved when using more qubits (i.e, wider quantum NNs), because it is sufficient to encode the principal features of handwritten digits with a few qubits.

\begin{figure}
	\centering
	\includegraphics[width=1.\columnwidth]{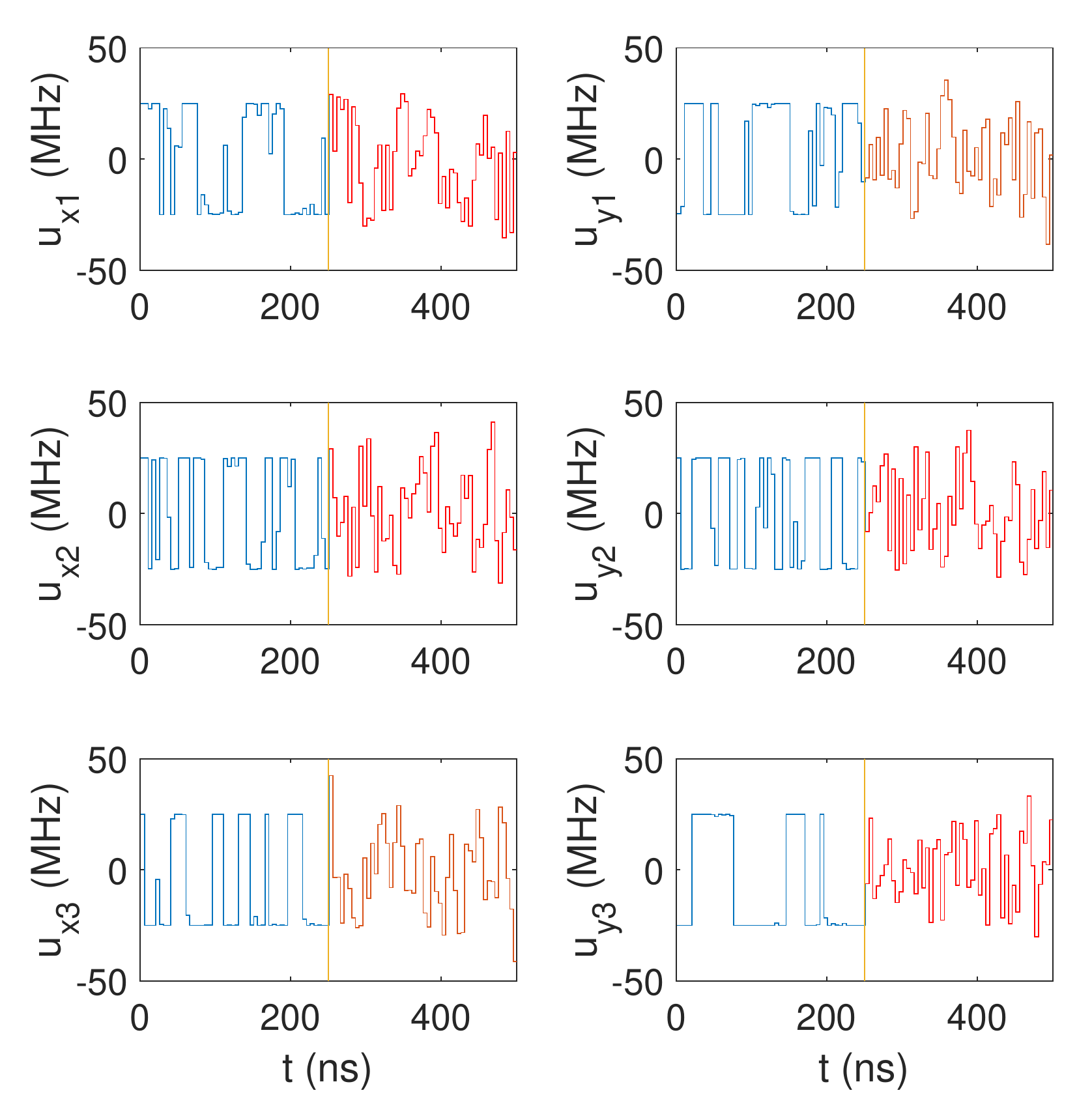}
	\caption{The control pulses applied in the 3-qubit end-to-end learning model with 50 encoding layers (blue color, $1\sim250$ns with 50 sampling periods) and 50 inference layers (red color, $251\sim500$ns with 50 sampling periods). }
	\label{fig:fields}
\end{figure}

Figure \ref{fig:fields} displays the optimized control pulses applied in a trained 3-qubit model with a randomly picked input sample. The first and second 50 sampling periods correspond to, respectively, the sample-dependent encoding layers (blue) and the sample-independent inference layers (red). Most encoding control variables reach the set bound $B=25$MHz. This pattern is observed in almost all simulations, implying that the encoding network may be further simplified (e.g., fix the control amplitudes and vary only the switching times) so as to reduce the model complexity. However, the saturated encoding control variables will lead to vanishing gradients along $W$ variables, which may slow down the training process on a landscape plateau \cite{Bottou2018}.

\subsection{The testing results}
The generalizability of the trained end-to-end learning models is tested with the validation dataset (containing 7837 independent samples belonging to the selected 8 classes). Table~\ref{tab:testing} lists both the empirical loss evaluated on the validation dataset and the error rates evaluated by an equivalent softmax function:
\begin{equation}
R_{\rm error} =P\left[\arg \max_{y}\langle \Psi_{x^{(j)}}({\bf w})| M_y| \Psi_{x^{(j)}}({\bf w})\rangle\neq y^{(j)}\right],
\end{equation}
which infers the label as $y$ for input $x$ if the probability of producing $y$ through the measurement is the largest. These indices are consistent with the learning curves tested on mini-batches, and are close to those of classical NN models. To our knowledge, such high performance (error rate lower than $10\%$) is only reachable with PQC-based models either on a downsize dataset (e.g., binary classification or with down-sampled images \cite{Grant2018,Farhi2018,Huggins2019}) or with more ($\geq9$) qubits \cite{Kerenidis2018,Wilson2018,Sweke2019,Henderson2020}.

\begin{table}[t]
\caption{The empirical losses and error rates evaluated on the validation MNIST dataset.}
\label{tab:testing}
\vskip 0.15in
\begin{center}
\begin{tabular}{c|c|c}
\hline
Model & ~~~~~~Loss~~~~~~ & Error Rate \\
\hline
3-QUBIT (10/10) & $13.82\%$ & $8.15\%$ \\
\hline
3-QUBIT (50/50) & $9.56\%$ & $3.30\%$ \\
\hline
4-QUBIT (10/10) & $13.78\%$ & $8.18\%$ \\
\hline
5-QUBIT (10/10) & $14.06\%$ & $8.80\%$ \\
\hline
\end{tabular}
\end{center}
\vskip -0.1in
\end{table}

The confusion matrix listed in Table \ref{tab:best} provides more details for the validation results on the trained 3-qubit model with 50 encoding and 50 inference layers. It can be seen that the digit ``0" has the highest precision, meaning that it is least probably to be misclassified as other digits. The digit ``4" has the highest and lowest recall rate, i.e., it is the least probable digit for other digits to be misclassified as. Among all the digits, ``0" is the relatively best recognized digit by the quantum learning model.

\begin{table}
 	\caption{The Confusion matrix, Precision ($\%$) and Recall Rates ($\%$) for the 8 digits classified by the 3-Qubit (50/50) quantum end-to-end learning model.}
 	\label{tab:best}
 	\begin{center}
 		\begin{tabular}{c|cccccccc|c}
 			\hline
	 & 0& 2&3&4&5&6&8&9 & Prec. \\
	\hline
    0 &     967 &     4  &    2    &    0  &   2 &   4  &   1  & 0    & 98.7  \\
    2 &      7  &  1000  &    9    &    3  &   0 &   1  &  10  & 2    & 96.9  \\
    3 &      0  &    11  &  982    &    0  &   4 &   0  &  11  & 2    & 97.2  \\
    4 &      0  &     4  &    0    &  953  &   0 &   9  &   1  & 15   & 97.1   \\
    5 &      5  &     0  &   14    &    0  & 851 &   8  &   9  & 5    & 95.4  \\
    6 &      4  &     2  &    0    &    2  &  14 & 931  &   5  & 0    & 97.2  \\
    8 &      5  &     8  &    8    &    6  &   5 &   4  & 935  & 3    & 96.0  \\
    9 &      5  &     2  &   15    &   11  &   3 &   0  &  14  & 959  & 95.0    \\
 			\hline
 Rec. & 97.4 &	97.0 &	95.3 &	97.7 &	96.8 &	97.3 &	94.8 &	97.3 & 96.7\\
 \hline
 		\end{tabular}
 	\end{center}
 \end{table}

\subsection{Roles of quantum and classical NN layers}
We carried out additional numerical experiments to further understand the respective roles of the classical and quantum layers played in the end-to-end learning process.

\begin{figure}
	\centering
	\includegraphics[width=1.0\columnwidth]{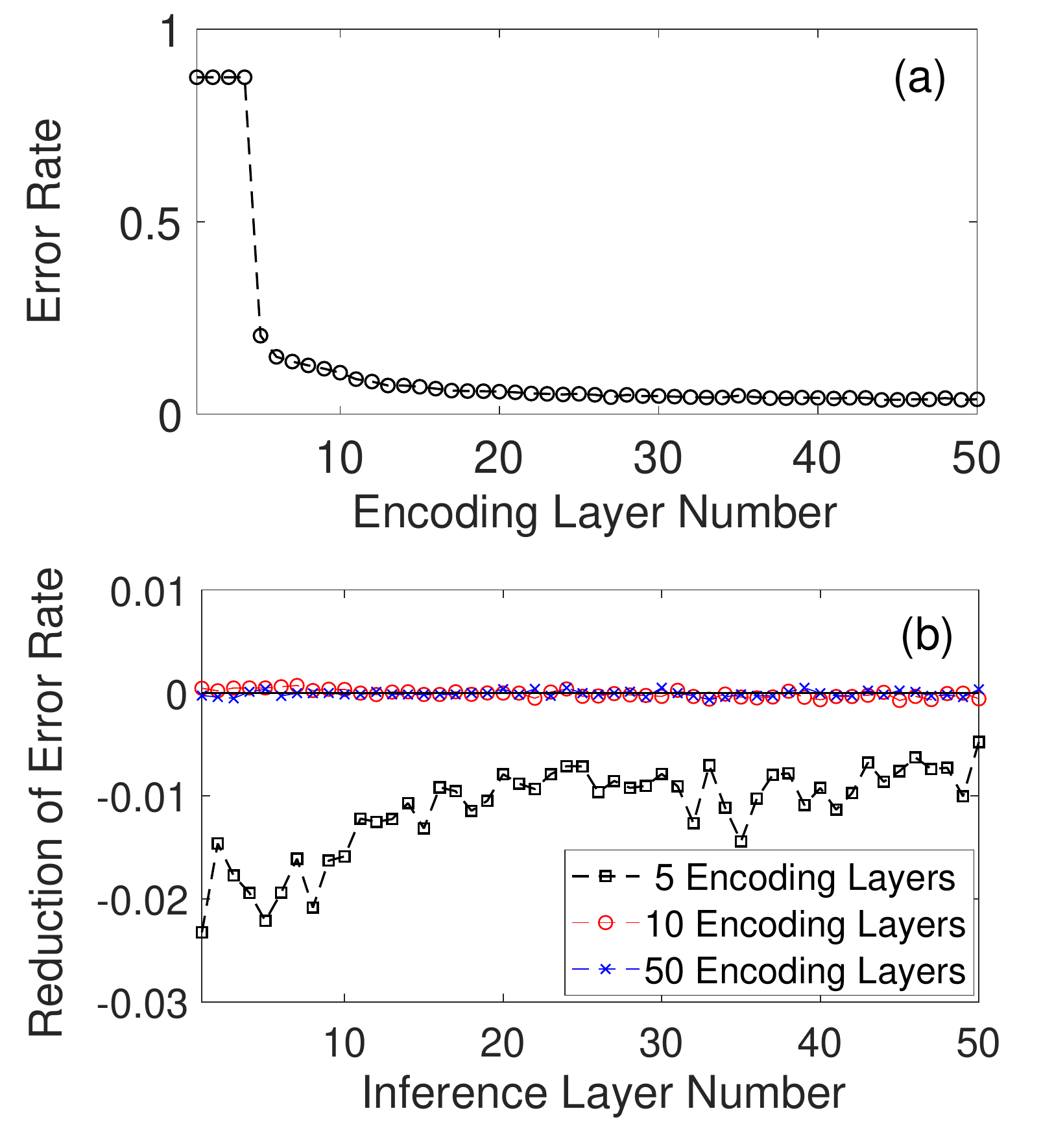}
	\caption{The simulated performance of the quantum end-to-end learning. (a) the error rates of 3-qubit models with $1\sim50$ encoding layers and no inference layers; (b) the error-rate reduction of 3-qubit models by inference layers based on pre-trained 5-layer, 10-layer and 50-layer encoding networks in (b), whose baseline error rates are $20.38\%$, $10.78\%$ and $3.87\%$, respectively.}
	\label{fig:comparison}
\end{figure}

We first randomly pick a $W$ matrix and fix it, and then train the inference layers. It turns out that the error rate can hardly be under $30\%$. However, if we remove the inference layers and train the encoding $W$ alone, the error rate [see Fig.~\ref{fig:comparison}(a)] can be easily reduced to about 10$\%$ with only 10 encoding layers. The performances can be further reduced using deeper encoding networks, but gradually saturates when the number of encoding layers is over 30.

{The above simulations imply that, with sufficiently deep encoding networks, the inference layers can be reduced to an identity mapping followed by measurements.} To see this more clearly, we fix the trained $W$ with 5, 10 and 50 encoding layers and train various numbers of inference layers. The results shown in Fig.~\ref{fig:comparison}(b) indicate that these additional inference layers do not significantly improve the performance except when there are only few encoding layers. This is consistent with the practice of classical deep learning in which feature selection is dominantly more important than the inference from the selected features.

{Because the size of the data-to-control interface linearly increases with the number of quantum encoding layers, it is necessary to verify whether the quantum network actually takes effect in the high performance achieved in Fig.~\ref{fig:comparison}(a). We follow the linear baseline rule \cite{Wilson2018} to specify the contributions of the quantum and classical layers by excluding their nonlinear effects that power up machine learning algorithms. In our model, the nonlinearity comes from the sigmoid function in the data-to-control interface and the control-to-state mapping. We first remove the sigmoid function from the data-to-control interface in the 3-qubit model with 50 encoding layers and no inference layers. With the remained control-to-state nonlinearity, the error rate is only increased from $3.87\%$ to $6.04\%$. However, if we further remove the control-to-state nonlinearity by replacing the encoding layers by a linear classical mapping, which forms a $784\times300\times3$ two-layer linear network, the achievable error rate rises much higher to $25.24\%$. These results clearly show the important role of the control-to-state nonlinearity associated with quantum encoding layers.}

\subsection{Robustness to noises}
The proposed quantum end-to-end learning model can also be trained to be robust to noises that are common on NISQ devices. The online training algorithm does not need to be changed because the system also receives ``samples" of the noises in addition to the samples fed from the dataset. In other words, the learning model is jointly trained by the dataset and the noises, from which the model gains simultaneously generalizability on the learning task and robustness to the noises \cite{Wu2019}.

\begin{figure}
	\centering
	\includegraphics[width=1.0\columnwidth]{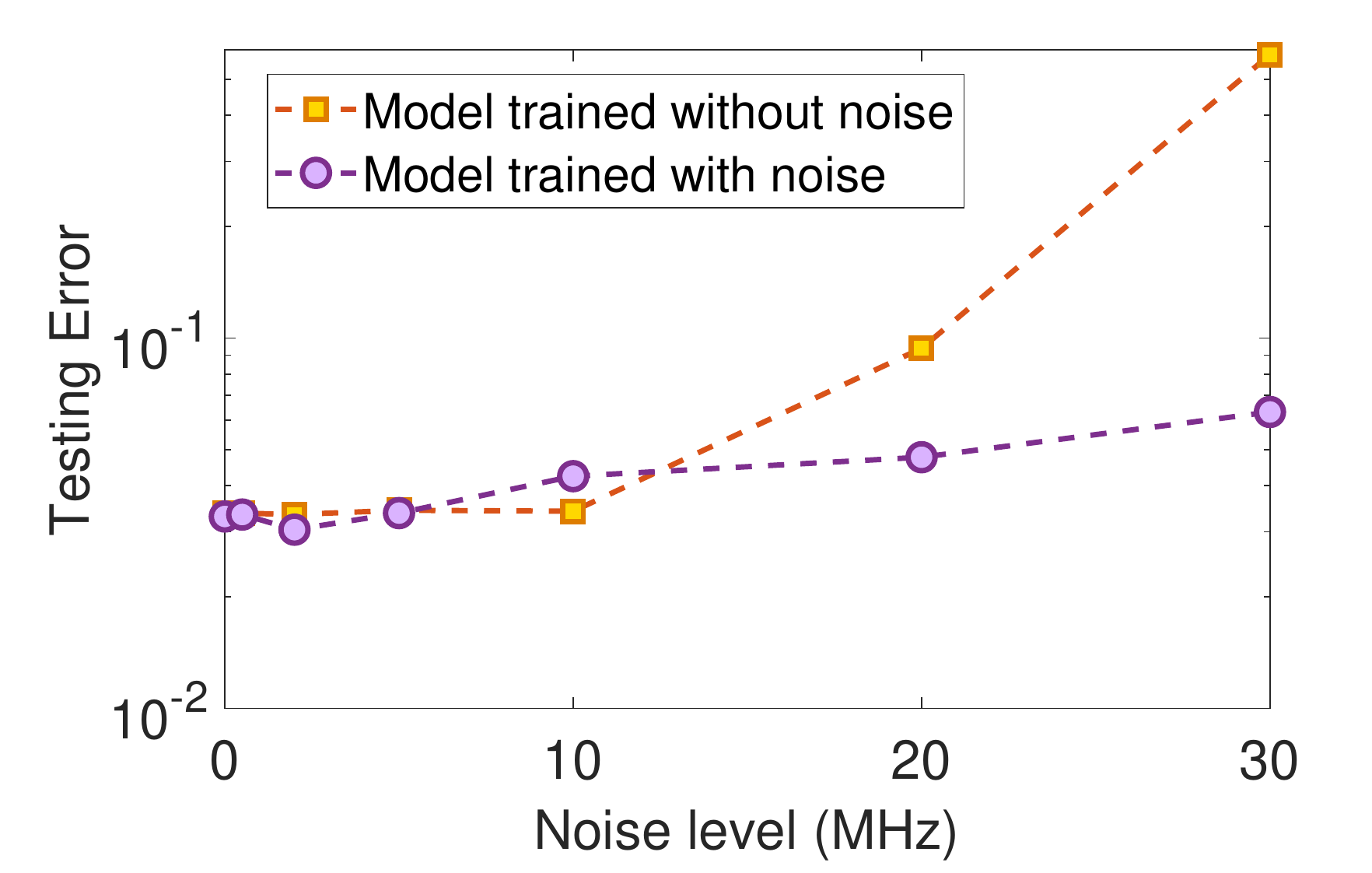}
	\caption{The testing errors of the end-to-end learning models trained with and without flux noises, respectively, in which the testing is done at various noise levels from $\delta=0$MHz to $\delta=30$MHz. The jointing training with both MNIST samples and flux noises can keep the error rate below $10\%$ under all simulated noise levels, while the training without noises leads to much worse performance when the noises are strong.}
	\label{fig:testnoise}
\end{figure}

For illustration, we add white flux noises $n_k(t)$ to the nomial three-qubit Hamiltonian $H(t)$ defined in Eq.~(\ref{eq:hamiltonian}) through the following Hamiltonian
\begin{equation}\label{}
  H_{\rm n}(t) = \sum_{k=1}^n n_k(t)\sigma_z^k.
\end{equation}
In the simulation, we train end-to-end learning models with 50 encoding and 50 inference layers at various noise levels (characterized by the variance $\delta$). Then, we compare their accuracies with that of the model trained without noise, where the testing is done with the same MNIST testing set and flux noises at the same level. As is shown in Fig.~\ref{fig:testnoise}, the accuracies of the two models are not very different when the noises are relatively weak ($\delta<10$MHz). When the noises are stronger, the error of the model trained with noises only slightly increases, while the model trained without noises performs much worse and becomes totally unreliable. The comparison shows that the practically trained learning model is inherently robust to noises.

\section{Concluding remarks}\label{sec:conclusion}
To summarize, we proposed a hardware-friendly quantum end-to-end learning model that can be conveniently deployed on NISQ processors. The model implements the controlled quantum dynamics as a quantum NN parameterized by experimental addressable control pulses, and the embedded data-to-control interface can automatically select appropriate features for inference. Numerical tests on the benchmarking MNIST dataset demonstrate that high performance can be attained with only a few qubits without downsizing the images even in presence of noises in the physical systems. Taking into accounts of the precision, the size of dataset, and the model complexity, the scheme achieves the best overall performance to our knowledge, exhibiting great potentials on real-world learning tasks.

Our proposal turns the machine learning process into an optimal control problem, both of which can be resolved with stochastic gradient-descent algorithms. This interesting connection can be dated back to the invention of famous BackPropagation (BP) algorithm for trainging NNs, which was derived from Pontryagin Maximum's Principle (PMP) in optimal control theory \cite{LeCun1988}. Recently, it was rediscovered to train deep NNs \cite{Li2017,Li2018}. From the opposite side, the design of robust quantum controls \cite{Wu2019} and quantum optimizers \cite{Dendukuri2019,Dong2019} can be taken as the design of a generalizable learning model. We expect to develop more efficient and noise-resilient training algorithms by unifying these two different but connected fields.

Viewing from the side of control theory, the capacity of the quantum end-to-end learning model can be partially understood through the controllability of the underlying control system (i.e., the ability of generating arbitrary unitary transformations), which is jointly determined by the physical qubit-qubit connectivity and the bandwidth/length of the applied control pulses. Full controllability is seemingly not required for many-qubit quantum processors, as quantum supremacy can be approached only with those transformations reachable in polynomial time \cite{Arenz2018}. However, we still suggest that the quantum hardware should be as controllable as possible, not only for larger model capacity but also for efficient search for high-performance learning models, because the underlying control landscape (for training process) tends to be free of traps  \cite{Rabitz2004,Wu2019a}.

Finally, we indicate that the proposed learning scheme can be easily extended to NISQ processors containing more qubits and other components (e.g., cavity modes or multi-level atoms), and the framework is also transplantable to any other learning tasks. In the simulation examples on the relatively simple MNIST dataset, the full power of the quantum end-to-end learning has not been fully released, and we expect to explore its potential power on more complicated learning tasks in future studies.

\acknowledgements
This work is supported by the National Key R$\&$D Program of China (Grants No.~2017YFA0304304 and No.~2018YFA0306703), NSFC (Grants No.~61833010 and No.~61773232), the Key-Area R$\&$D Program of GuangDong Province (Grant No. 2018B030326001) and a grant from the Institute for Guo Qiang, Tsinghua University. Invaluable discussions with Prof. Changshui Zhang are greatly appreciated.

\end{document}